\documentclass[twocolumn,showpacs,preprintnumbers,amsmath,amssymb,nofootinbib]{revtex4}

\usepackage{graphicx}
\usepackage{color}
\usepackage{dcolumn}
\usepackage{bm}
\newcommand{\ps}{p\hspace{-0.44em}/\hspace{0.06em}}
\newcommand{\eq}[1]{Eq.~(\ref{#1})}
\newcommand{\eqsand}[2]{Eqs.~(\ref{#1}) and (\ref{#2})}

\newcommand{\dd}{\ensuremath{D\!-\!\Dbar{}\,}}
\newcommand{\kk}{\ensuremath{K\!-\!\Kbar{}\,}}
\newcommand{\bbms}{\bbs\ mixing}
\newcommand{\bbmd}{\bbd\ mixing}

\newcommand{\ddm}{\dd\ mixing}
\newcommand{\kkm}{\kk\ mixing}
\newcommand{\bbd}{\ensuremath{B_d\!-\!\Bbar{}_d\,}}
\newcommand{\bbs}{\ensuremath{B_s\!-\!\Bbar{}_s\,}}

\newcommand{\Bbar}{\,\overline{\!B}}
\newcommand{\Dbar}{\,\overline{\!D}}
\newcommand{\Kbar}{\,\overline{\!K}}
\newcommand{\real}{\mathrm{Re}\,}

\newcommand{\gev}{\, \mathrm{GeV}}
\newcommand{\tev}{\, \mathrm{TeV}}

\begin{document}
\linespread{1.1}
\preprint{TTP11-14}

\title{Phenomenological consequences of radiative flavor violation in the MSSM}
%
\author{Andreas Crivellin${}^{1,2}$, Lars Hofer${}^{1,3}$, Ulrich Nierste${}^{1}$ and Dominik Scherer${}^{1}$\bigskip}

\affiliation{${}^{1}$Institut f\"ur Theoretische Teilchenphysik\\
               Karlsruhe Institute of Technology, 
               Universit\"at Karlsruhe \\ D-76128 Karlsruhe, Germany\bigskip\\
               ${}^{2}$Albert Einstein Center for Fundamental Physics, Institute for Theoretical Physics,\\
               Universit\"at Bern, CH-3012 Bern, Switzerland\bigskip\\
              ${}^{3}$Institut f\"ur Theoretische Physik und Astrophysik\\ 
               Universit\"at W\"urzburg \\ D-97074 W\"urzburg, Germany\bigskip}

\begin{abstract}
  In this article we investigate the consequences of radiative flavor
  violation (RFV) in the Minimal Supersymmetric Standard Model (MSSM).
  In this framework the small off-diagonal elements of the
  Cabibbo-Kobayashi-Maskawa (CKM) matrix and the small quark masses of
  the first two generations are generated from the trilinear
  supersymmetry (SUSY)-breaking terms. The impact of RFV on flavor-physics
  observables is studied in detail. We focus on the limiting cases in
  which the CKM matrix is either generated in the down-sector, i.e. by
  the soft SUSY-breaking mass insertions $\delta^{d\,LR}_{i3}$
  ($i=1,2$), or in the up-quark sector, i.e.\ by the mass insertions
  $\delta^{u\,LR}_{i3}$. In the first case we find an enhancement of
  $b\to s\gamma$, which constrains the allowed range of sparticle
    masses (Fig.~\ref{b-s-gamma-allowed}). In addition, neutral Higgs
  penguins significantly contribute to $B_{s,d}\to \mu^+\mu^-$ and, if
  also $\delta^{d\;LR}_{32}$ is different from zero, these Higgs effects
  are capable of explaining the observed CP phase in the $B_s$ system.
  If, on the other hand, the CKM generation takes place in the
  up-sector, $|\epsilon_K|$ receives additional positive contributions
  enforcing large squark and gluino masses (see
  Fig.~\ref{K-allowed}). In this case also the rare decay $K\to
  \pi\nu\bar{\nu}$ receives sizable contributions. In conclusion we find
  that for SUSY masses around 1~$\tev$ RFV is an interesting alternative
  to Minimal Flavor Violation (MFV).
\end{abstract}

\maketitle                             
\section{Introduction}
The smallness of the fermion masses of the first two generations and of
the off-diagonal CKM elements suggest the idea that these quantities are
perturbations, induced by quantum loop corrections.  Already in 1972
S.~Weinberg explored this idea \cite{Weinberg:1972ws}.  Later this
possibility has been studied in the contexts of Grand Unified Theories
\cite{Ibanez:1981nw} and mirror families
\cite{Kagan:1987wf,Kagan:1989fp}. In 1982 and 1983 several authors
realized that the trilinear soft supersymmetry (SUSY)-breaking terms
\cite{Donoghue:1983mx} can indeed generate fermion masses radiatively in
the MSSM
\cite{Lahanas:1982et,Buchmuller:1982ye,Nanopoulos:1982zm,Masiero:1983ph}.
This possibility was later worked out in more detail by T.\ Banks
\cite{Banks:1987iu}. (A review of different mechanisms of radiative mass
generation can be found in \cite{Babu:1989fg}.) The fermion masses and
off-diagonal CKM elements can be viewed to arise from \emph{soft Yukawa
  couplings}, generated through loops involving a trilinear term
$A_{ij}^q$, with $q=u,d$ and $i,j=1,2,3$ labeling the fermion
generation. While the usual hard Yukawa couplings in the superpotential
are identical in Higgs and higgsino couplings to quarks and squarks, the
corresponding soft Yukawa couplings are very different from each other.
An extensive study of soft Yukawa couplings in supersymmetric theories
was carried out in 1999 by F.\ Borzumati et
al.~\cite{Borzumati:1999sp}. An alternative possibility to generate CKM
elements radiatively arises in left-right symmetric models
\cite{Babu:1998tm}. In 2004 SUSY-breaking scenarios which can give rise
to radiative masses have been studied \cite{Ferrandis:2004ri}.  The
required non-minimal flavor structure of $A_{ij}^q$ unavoidably affects
flavor-changing neutral current (FCNC) processes. The precision studies
of FCNC at B-factories and the Tevatron in the past decade therefore
challenge the idea of RFV. However, recently two of us revisited RFV in
Refs.~\cite{Crivellin:2008mq,Crivellin:2009pa} and found that it is
possible to generate each element of the CKM matrix separately without
violating bounds from FCNC processes for squark masses above
approximately $500\,\rm{GeV}$.

The framework of Refs.~\cite{Crivellin:2008mq,Crivellin:2009pa} is
  as follows: The $[U(3)]^3$ flavor symmetry of the gauge sector (we
  neglect leptons here) is broken to $[U(2)]^3\times U(1)$ by non-zero
  hard Yukawa couplings $y_{b,t}\neq 0$ for bottom and top quarks,
  respectively.  Then either $A_{ij}^u$ or $A_{ij}^d$ is chosen as the
  spurion breaking $[U(2)]^3\times U(1)$ to $U(1)_B$, the baryon
  number symmetry. The bilinear SUSY-breaking terms are chosen
  universal, i.e. they respect the $[U(3)]^3$ symmetry of the gauge
  sector, up to renormalization effects from the $A_{ij}^q$. Our
model of radiative mass and CKM generation has several advantages
compared to the generic MSSM:
\begin{itemize}
        \item The imposed $[U(2)]^3$ symmetry of the Yukawa sector protects
          the quarks of the first two generations from a tree-level
          mass term. The smallness of their masses is thus explained
          naturally via loop-suppression.
	\item The model is economical: Flavor violation and SUSY
          breaking have the same origin.
	\item Flavor universality holds for the first two
          generations. Thus our model is minimally flavor-violating
          with respect to the first two generations since the quark
          and the squark mass matrices are diagonal in the same
          basis \footnote{Note that our definition of MFV differs from
            the one of Ref.~\cite{D'Ambrosio:2002ex} in the sense that
            we refer to the effective Yukawa couplings and not
            the hard couplings of the superpotential.};
            i.e. the off-diagonal elements $\Delta_{12}^{q\,XY}$,
              $X,Y=L,R$, of the squark mass matrices vanish in the
              basis of the superfields in which the quark mass
              matrices are diagonal.  This explains why $K$ and
              $D$ physics data comply well with the Standard Model
              predictions. However, double mass insertions involving
            the third generation affect the transitions between the
            first two generations (see section~\ref{sec:FCNC} for
            details) permitting small deviations from the CKM pattern.
	\item The SUSY flavor problem is reduced to the quantities
          $\delta^{q\,RL}_{13,23}$ ($\delta^{q\;XY}_{ij}=\Delta^{q\;XY}_{ij}/m_{\tilde q}^2$ with 
          $m_{\tilde q}$ denoting the average squark mass) 
          because they are the only
          flavor-changing SUSY-breaking terms which are not related to
          corresponding CKM elements. However, these parameters are
          less constrained from FCNCs than $\delta^{q\,LR}_{13,23}$ 
          and can explain a potential new CP phase
          indicated by recent data on $B_s$ mixing
          \cite{Lenz:2010gu}, as we will show below. 
          Furthermore, as shown in
          Ref.~\cite{Crivellin:2009sd} $\delta^{u\,RL}_{13,23}$ can
          also induce a right-handed W coupling which can explain
          discrepancies between inclusive and exclusive determinations
          of $V_{ub}$ and $V_{cb}$.
	\item The SUSY CP problem is substantially alleviated by an
          automatic phase alignment~\cite{Borzumati:1999sp} between
          the $A$-terms and the effective Yukawa couplings. In
          addition, the phase of $\mu$ essentially does not enter the
          electric dipole moments (EDMs) of the light quarks at the one-loop level because the
          Yukawa couplings of the first two generations are zero.
	\item   When our RFV framework is extended to leptons,
          the anomalous magnetic moment of the muon receives a
          contribution (independent of $\mu$) which interferes
          constructively with the SM one. In this way the discrepancy
          between the SM prediction and experiment can be
          solved~\cite{Borzumati:1999sp,AJU}.
\end{itemize}

In this article we investigate the implications of this model of RFV for flavor-changing processes in the quark sector and we complement the analyses of Refs.~\cite{Crivellin:2008mq,Crivellin:2009pa} in some important points:
\begin{itemize}
	\item As shown in Ref.~\cite{Crivellin:2009ar}, chirally enhanced corrections to FCNC processes are important in the presence of flavor-changing $A$-terms. Therefore, we include these sizable corrections into our analysis.
	\item Diagrams involving the double mass insertions $\delta^{q\;LR}_{13}\delta^{q\;LR}_{23}$ which contribute to $K$--$\overline{K}$ and $D$--$\overline{D}$ mixing are considered.
	\item Taking into account chirally enhanced corrections and multiple flavor changes  we explicitly show the allowed regions in parameter space in the $m_{\tilde g}-m_{\tilde q}$ plane. 
	\item Predictions for the rare decay $K\to\pi\nu\bar{\nu}$ are given.
	\item Off-diagonal $A$-terms induce (non-decoupling) flavor-changing neutral Higgs couplings proportional to $\tan\beta$~\cite{Crivellin:2010er}. We study the effect of these couplings on $B_s\to\mu^+\mu^-$ and $B_s$-$\bar{B}_s$ mixing. 
\end{itemize}

\section{Radiative mass and CKM Generation}

As discussed in the introduction, the light quark masses (possibly also
the bottom mass) and the off-diagonal CKM elements can be induced in the
MSSM via self-energy diagrams involving the trilinear $A$-terms. These
self-energies are chirally enhanced gaugino-sfermion loops which modify
the relations between physical masses and Yukawa couplings significantly
\footnote{By using 't Hooft's naturalness argument very strong bounds on
  the mass-insertions $\delta^{f\;LR}_{11,22}$ can be derived by
  requiring that the supersymmetric corrections do not exceed the
  measured masses \cite{Crivellin:2008mq,Crivellin:2010gw}}. In this
section we define our framework and quantify the size of the $A$-terms
needed to generate the masses and the off-diagonal CKM elements
radiatively.\medskip

While in the MSSM the light fermion masses can arise from
loop-induced Higgs couplings involving virtual squarks and
gluinos \footnote{Of course also the bino diagram contributes to the
  quark masses, but it is suppressed by a factor
  $\frac{3\alpha_1}{8\alpha_s}$}, the heaviness of the top quark
requires a special treatment for $Y^t$. The successful bottom-tau
Yukawa unification suggests to keep the tree-level Yukawa couplings
for the third generation lepton and down-type quark, as well.  At
large $\tan \beta$, this idea gets even more support from the
successful unification of the top and bottom Yukawa couplings, as
suggested by some SO(10) GUTs.  In the modern language of
Refs.~\cite{D'Ambrosio:2002ex,Chivukula:1987py} the global $[U(3)]^5$
flavor symmetry of the gauge sector is broken down to $[U(2)]^5 \times
[U(1)]^2$ by the Yukawa couplings of the third generation. Here the
five $U(2)$ factors correspond to rotations of the left-handed
doublets and the right-handed singlets of the first two generation
quarks and leptons in flavor space, respectively.  Imposing this
symmetry on the Yukawa sector implies
\begin{equation}
Y^{f(0)}  = \left( {\begin{array}{*{20}c}
   0 & 0 & 0  \\
   0 & 0 & 0  \\
   0 & 0 & y^f  \\
\end{array}} \right),\;\;\;V^{(0)}  = \left( {\begin{array}{*{20}c}
   1 & 0 & 0  \\
   0 & 1 & 0  \\
   0 & 0 & 1  \\
\end{array}} \right)\label{eq:YukCKM}
\end{equation}
in the bare Lagrangian. The absence of tree-level Yukawa couplings of the light fermions as well as of off-diagonal CKM elements requires that these quantities have to be generated via radiative corrections.\medskip

While the trilinear SUSY-breaking terms $A^u$ and $A^d$ are the spurions
breaking the $[U(2)]^3\times U(1)$ symmetry of the hard quark Yukawa sector, we assume that the bilinear squark mass terms
${\bf{M}}_{\tilde q\;LL,RR}$ possess the full $[U(3)]^3$ flavor symmetry
of the (s)quark sector at some high scale.  At the low electroweak scale
the $[U(3)]^3$ symmetry of ${\bf{M}}_{\tilde q\;LL,RR}$ is broken by two
renormalisation effects: First, there are renormalization-group (RG)
effects proportional to Yukawa couplings. These effects split the third
eigenvalue from the first two ones, but do not induce off-diagonal terms
in ${\bf{M}}_{\tilde q\;LL,RR}$. Second, the trilinear terms $A^u$ and
$A^d$ will also renormalize ${\bf{M}}_{\tilde q\;LL,RR}$. The
off-diagonal terms induced by the RG evolution lead to flavor-changing
elements in ${\bf{M}}_{\tilde q\;LL,RR}$. In a given FCNC loop diagram
these terms lead to additional contributions, which, however, are
governed by the same element of $A^q$ which enters the considered loop
directly. Therefore the RG effects in ${\bf{M}}_{\tilde q\;LL,RR}$
simply shift the numerical value of $A^q_{jk}$ determined from the
condition that RFV reproduces the measured CKM elements. In our
phenomenological study, treating the trilinear terms as independent, one
can therefore neglect the RG effects in ${\bf{M}}_{\tilde q\;LL,RR}$.
The situation is different, once GUT boundary conditions are placed on
the RFV model.  \medskip

Let us first have a look at the consequences of this symmetry-breaking
pattern for the first two quark generations. Here we can exploit
the $SU(2)$ flavor symmetry of the gauge- and Yukawa-sector to choose
a basis for the left- and right-handed quark fields in which for
example the upper left $2\times2$ block of $A^d$, denoted by
$A^d_{2\times 2}$, is diagonal. 
Choosing four different rotations for left- and right-handed
up and down quark fields {(so that the $SU(2)_L$ gauge
  symmetry is no more manifest)}, we can even diagonalize
$A^d_{2\times 2}$ and $A^u_{2\times 2}$ simultaneously. In this step
the tree-level Cabibbo matrix $(V^{(0)}_C)_{2\times 2}$ is
generated. Since in this basis no sources of flavor-violating (1,2)
elements are present in the squark-mass matrices, the so-obtained 
$(V^{(0)}_C)_{2\times 2}$ equals the Cabibbo matrix $V_{2\times 2}$
known from experiment (up to negligible corrections arising from loops
involving a $1\to3\to 2$ transition). This observation implies that
$A^q_{2\times 2}$ is proportional to the corresponding effective
(loop-induced) Yukawa matrix $(Y^q_{\textrm{eff}})_{2\times 2}$.  Note
that even with respect to the first two generations the model is not
minimally flavor-violating in the literal sense of
Ref.~\cite{D'Ambrosio:2002ex}: The $A$-terms cannot be constructed out
of the (vanishing) tree-level Yukawa couplings and vice versa.
However, our model obeys the MFV definition with respect to the first
two generations, if one defines MFV with
$(Y^q_{\textrm{eff}})_{2\times 2}$ instead, and the Cabibbo matrix is
the only source of flavor violation.
\medskip
  
For the third generation the situation is different. The direction of
the third generation in flavor space is already fixed from the Yukawa
sector, by the requirement of diagonal Yukawa matrices
$Y^{q(0)}$. Therefore, the elements $A^q_{3i}$ and $A^q_{i3}$
($i=1,2$) cannot be eliminated by a redefinition of the flavor
basis. The effect of these $A$-terms is twofold: On one hand, they
have to generate the effective CKM elements $V_{3i}$ and $V_{i3}$. On
the other hand, they act as sources of non-minimal flavor violation
and thus they are constrained from FCNC processes.\medskip

It is common to choose a basis for the quark fields in which the
Yukawa couplings are diagonal and, in order to have manifest
supersymmetry in the superpotential, to subject the squarks to the
same rotations as the quarks. The resulting basis for the super-fields
is called the super-CKM basis. However, since the Yukawa couplings of
the first two generations are zero in our scenario, the super-CKM
basis is not defined unambiguously. We {fix} this ambiguity with 
the additional requirement of diagonal $A^d_{2\times 2}$ and $A^u_{2\times
  2}$ (as described above). The (bare) CKM matrix and the $A$-terms
then take the following form:
\begin{eqnarray}
V^{(0)}_C  &=& \left( {\begin{array}{*{20}c}
   \cos\theta_C &   \sin\theta_C & 0  \\
     -\sin\theta_C &   \cos\theta_C & 0  \\
   0 & 0 & 1  \\
\end{array}} \right),\label{CKM_Cabbibo}\\[0.2cm]
A^q_C &=& \left( {\begin{array}{*{20}c}
   A^q_{11}&   0 & A^q_{13}  \\
   0 &   A^q_{22} & A^q_{23}  \\
   A^q_{31} & A^q_{32} & A^q_{33}  \\
\end{array}} \right).\label{cnice}
\end{eqnarray}
The subscript $C$ denotes the Cabbibo rotation performed for this
choice of the super-CKM basis.  The Cabbibo angle stems from the
misalignment between the $A^u$-terms and the $A^d$-terms of the first
two generations in any weak basis.  In the basis
  corresponding to \eqsand{CKM_Cabbibo}{cnice}, however, $A^d_{2\times
    2}$ and $A^u_{2\times 2}$ are simultaneously diagonal and the
  familiar Cabibbo angle $\theta_C$ appears in the $W$ couplings to
  (s)quarks.\medskip

Since the bare Yukawa couplings of quarks of the first two
generations are zero, these quarks do not develop tree-level mass
terms. Because of the non-vanishing $A$-terms of the first two
generations they can, however, couple to the Higgs fields and to
their vacuum expectation values (vevs) via a loop with SUSY-particles. The corresponding self-energy
diagrams generate effective masses for the quarks.\medskip

\begin{figure*}
\centering
\includegraphics[width=0.75\textwidth]{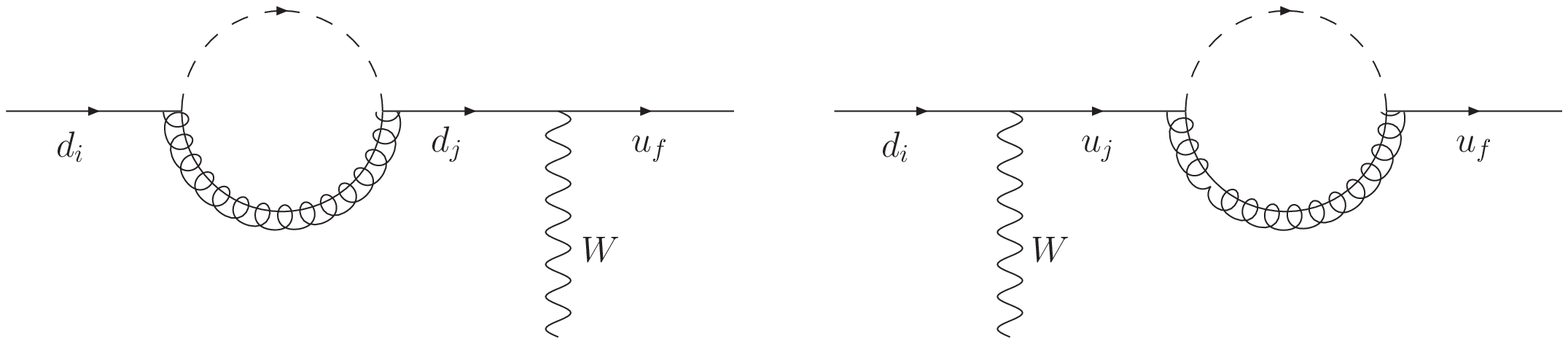}
\caption{One-loop contributions to the CKM matrix from the down-sector and from the up-sector, respectively. \label{W-Diagramm}}\hrule
\end{figure*}
At this point we recall some results concerning quark self-energies,
the renormalization of masses and flavor-valued {field rotations}. It
is possible to decompose any self-energy into its chirality-flipping
and its chirality-conserving parts as
\begin{eqnarray}
\Sigma _{ij}^q(p)  &=& \left( {\Sigma _{ij}^{q\;LR}(p^2)  + \ps\Sigma _{ij}^{q\;RR}(p^2) } \right)P_R 
\nonumber\\ &&+ \left( {\Sigma _{ij}^{q\;RL}(p^2) + \ps\Sigma _{ij}^{q\;LL}(p^2) } \right)P_L\,.
\label{self-energy-decomposition}
\end{eqnarray}
Only the chirality-flipping self-energies
$\Sigma^{q\;RL}_{ij}=\Sigma^{q\;LR*}_{ji}$ are capable of generating
sizable effective mass terms in the absence of tree-level Yukawas. Since
we are working with quarks we can concentrate on the contributions from
gluino-squark loops.\footnote{This is a good approximation for the
  self-energies. However, when we later consider FCNC processes also the
  chargino contributions are important. In principle also chirally enhanced chargino
  self-energies (in the down-quark sector) contribute to the CKM
  renormalization if~$A_{13,23}^{u}$ are unequal to zero. Even though
  these contributions are enhanced by a factor $m_t/m_b$ they are small
  due to a suppression by $v^2/M_{\rm{SUSY}}^2\times g_2^2/g_s^2$,
  especially at smaller values of $\tan\beta$.}  At vanishing external
momentum, the SUSY-QCD self-energy is given by (the conventions are defined
in the appendix of Ref.~\cite{Crivellin:2008mq})
\begin{equation}
\Sigma_{ij}^{q\;LR}= \dfrac{2 \alpha_s}{3 \pi}m_{\tilde{g}}\sum_{s = 1}^6 W^{\tilde q}_{is} W^{\tilde q *}_{(j+3)s} B_{0}(m_{\tilde{g}}^{2},m_{\tilde{q}_s}^{2}).
 \end{equation}
At first order in the mass insertion approximation this simplifies to
\begin{equation}
\Sigma_{ij}^{q\;LR} = \dfrac{2\alpha_s}{3 \pi} m_{\tilde{g}} \,\Delta^{q\,LR}_{ij}\, C_0\left(m_{\tilde{g}}^2, m_{\tilde{q}_{iL}}^2,m_{\tilde{q}_{jR}}^2 \right)\label{SQCD-SE-MIA}
\end{equation}
with
\begin{eqnarray}
\Delta^{u\,LR}_{ij}&=&-v_u A^u_{ij}\;-\;v_d\,\mu\, Y^{u(0)}_{ij}\,,\nonumber\\
\Delta^{d\,LR}_{ij}&=&-v_d A^d_{ij}\;-\;v_u\, \mu\, Y^{d(0)}_{ij}\,.\label{DeltaLR}
\end{eqnarray}
\medskip

Now we turn to the renormalization of quark masses and to the rotations in flavor-space which are induced by the self-energies. 
Including chirally enhanced corrections, the physical masses $m_{q_i}$, which are extracted from experiment using the SM prescription with ordinary QCD corrections renormalized in the $\overline{\rm{MS}}$ scheme, are given as
\begin{equation}
m_{q_i}\,=\,m_{q_3}^{(0)}\,\delta_{i3}  \,+\, \Sigma _{ii}^{q\;LR}.
\label{massrenormalization}
\end{equation}
Here $m^{(0)}_{q_3}=y^{q}v_q$ denotes the bare quark mass generated by the Yukawa coupling $y^q$ defined in \eq{eq:YukCKM}. Since 
$\Sigma _{ii}^{q\;LR}$ is finite, there is no need to renormalize $m^{(0)}_{q_3}$ by splitting it into a renormalized part 
and into a counter-term. For the first two quark generations, we directly read off the requirement of radiative mass generation 
from \eq{massrenormalization}:
\begin{equation}
\Sigma _{ii}^{q\;LR} = m_{q_i}\hspace{1cm} (i=1,2).
\label{radiative_M}
\end{equation}\medskip

The flavor-changing self-energies $\Sigma^{q\,LR}_{ij}$ induce 
{field} rotations
\begin{equation}
   \psi^{q\,L}_{i}\,\longrightarrow\, U^{q\,L}_{ij}\,\psi^{q\,L}_j\,
\end{equation}
in flavor space. To leading order in small quark mass ratios $U^{q\,L}$ reads
\begin{equation}
U^{q\,L}  = \begin{pmatrix}
1 & \dfrac{\Sigma^{q\,LR}_{12}}{m_{q_2}} & \dfrac{\Sigma^{q\,LR}_{13}}{m_{q_3}} \\[0.4cm]
-\dfrac{\Sigma^{q\,RL}_{21}}{m_{q_2}} & 1 & \dfrac{\Sigma^{q\,LR}_{23}}{m_{q_3}} \\[0.4cm]
-\dfrac{\Sigma^{q\,RL}_{31}}{m_{q_3}}\,+\,\dfrac{\Sigma^{q\,RL}_{21}}{m_{q_2}}\,
 \dfrac{\Sigma^{q\,RL}_{32}}{m_{q_3}} & 
-\dfrac{\Sigma^{q\,RL}_{32}}{m_{q_3}} & 1 \\
\end{pmatrix}.
\label{DeltaU}
\end{equation}
Note that the quark masses $m_{q_i}$ appearing in this equation are the physical $\overline{\textrm{MS}}$\,-\,masses which have to be evaluated at the same scale as the self-energies \cite{Crivellin:2008mq,Hofer:2009xb} and that $U^{q\,L}_{31}$ contains a two-loop contribution which can be numerically important \cite{Crivellin:2010gw}. The formula for $U^{q\,L}$ given in \eq{DeltaU} is valid irrespective of the basis chosen for the (s)quark superfields. In our super-CKM basis (with diagonal $A^d_{2\times 2}$ and $A^u_{2\times 2}$), $\Sigma^{q\,LR}_{12}=\Sigma^{q\,RL*}_{21}$ vanishes and the corresponding terms in $U^{q\,L}$ are absent:
\begin{equation}
U^{q\,L}_{C}  = \begin{pmatrix}
1 & 0 & \dfrac{\Sigma^{q\,LR}_{13}}{m_{q_3}} \\[0.4cm]
0 & 1 & \dfrac{\Sigma^{q\,LR}_{23}}{m_{q_3}} \\[0.4cm]
-\dfrac{\Sigma^{q\,RL}_{31}}{m_{q_3}} & 
-\dfrac{\Sigma^{q\,RL}_{32}}{m_{q_3}} & 1 \\
\end{pmatrix}.
\label{DeltaU_Cabbibo}
\end{equation}\medskip

\begin{figure*}
\centering
\includegraphics[width=0.45\textwidth]{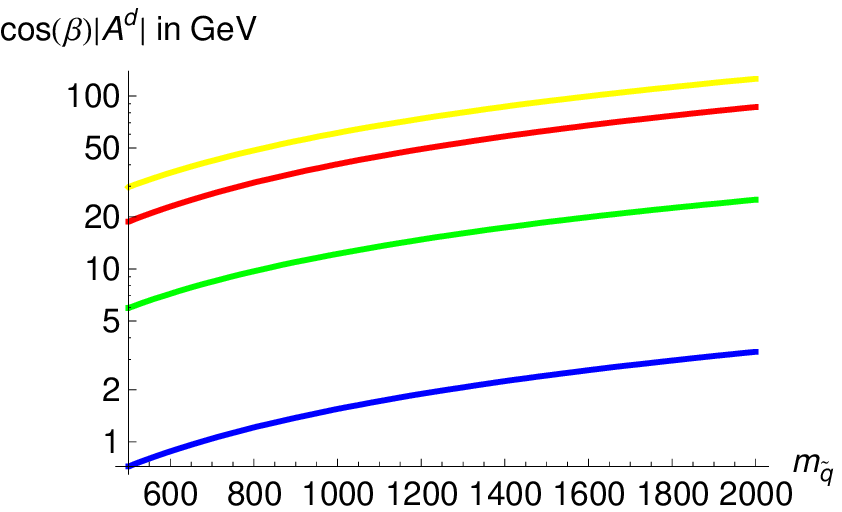}\hspace{0.08\textwidth}
\includegraphics[width=0.45\textwidth]{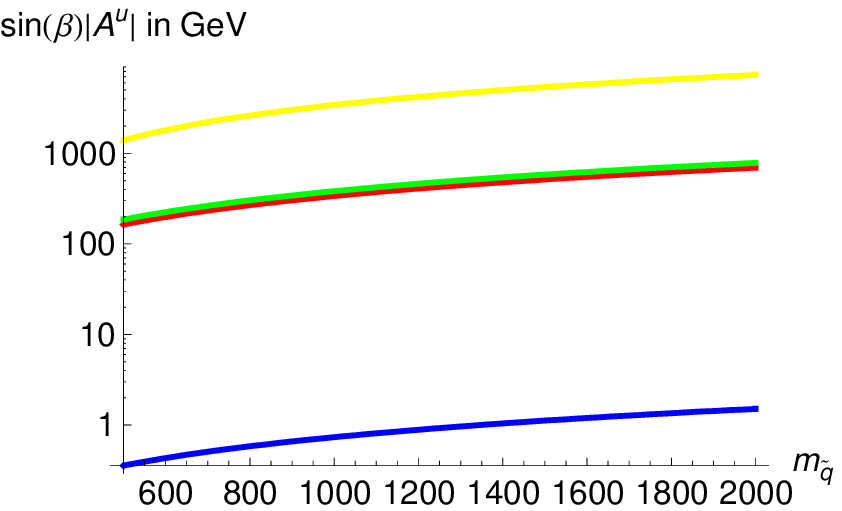}
\caption{Left: Absolute values of $A^d_{ij}$ (at scale
  $M_{\rm{SUSY}}$) needed to generate the down and strange
  quark masses and the off-diagonal CKM elements involving the third
  generation. Blue(darkest):~$A^d_{11}$, red:~$A^d_{22}$,
  green:~$A^d_{13}$, yellow(lightest):~$A^d_{23}$.  \newline Right:
  Absolute values of $A^u_{ij}$ (at scale $M_{\rm{SUSY}}$) needed to
  generate the up and the charm quark mass and the off-diagonal
  CKM elements involving the third generation. blue(darkest):
  $A^u_{11}$, red:~$A^u_{22}$, green:~$A^u_{13}$,
  yellow(lightest):~$A^u_{23}$. \label{Ad}}
\end{figure*}

The flavor-changing self-energies $\Sigma^{q\,LR}_{ij}$ induce
corrections to the CKM matrix $V^{(0)}_C$ as depicted in
Fig.~\ref{W-Diagramm}. In this way they generate the physical CKM
matrix $V$, measured in low-energy experiments. In terms of the
{field} rotations $U^{q\,L}_C$ the physical CKM matrix $V$ can be
expressed as
\begin{equation}
 V\,=\,U^{u\,L\dagger}_C\, V^{(0)}_C\, U^{d\,L}_C\,.
 \label{eq:CKMreno}
\end{equation}
\begin{figure*}
\centering
\includegraphics[width=0.45\textwidth]{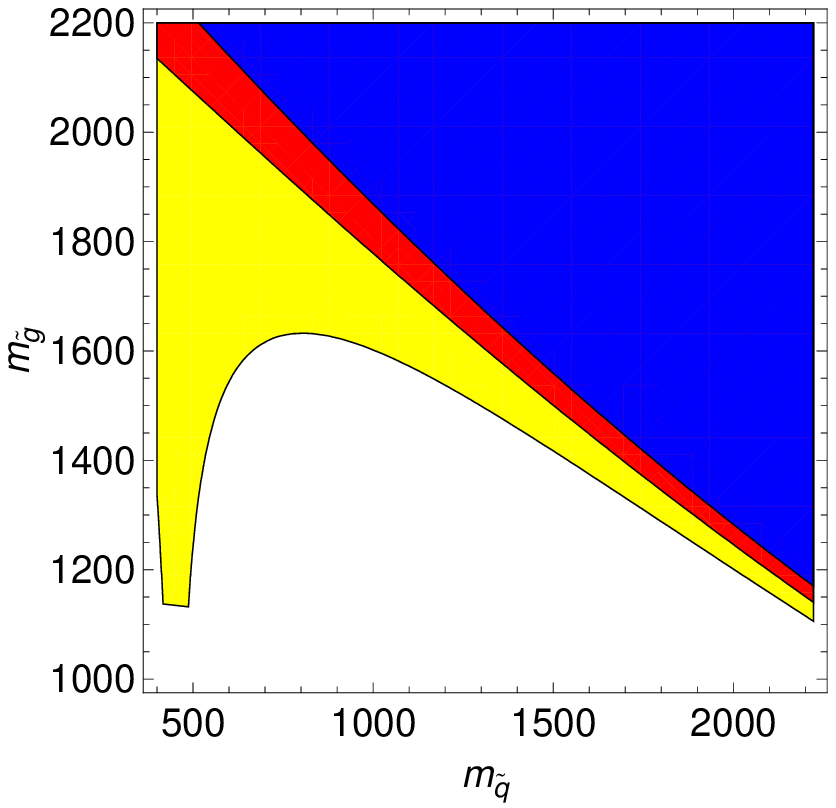}\hspace{0.07\textwidth}
\includegraphics[width=0.47\textwidth]{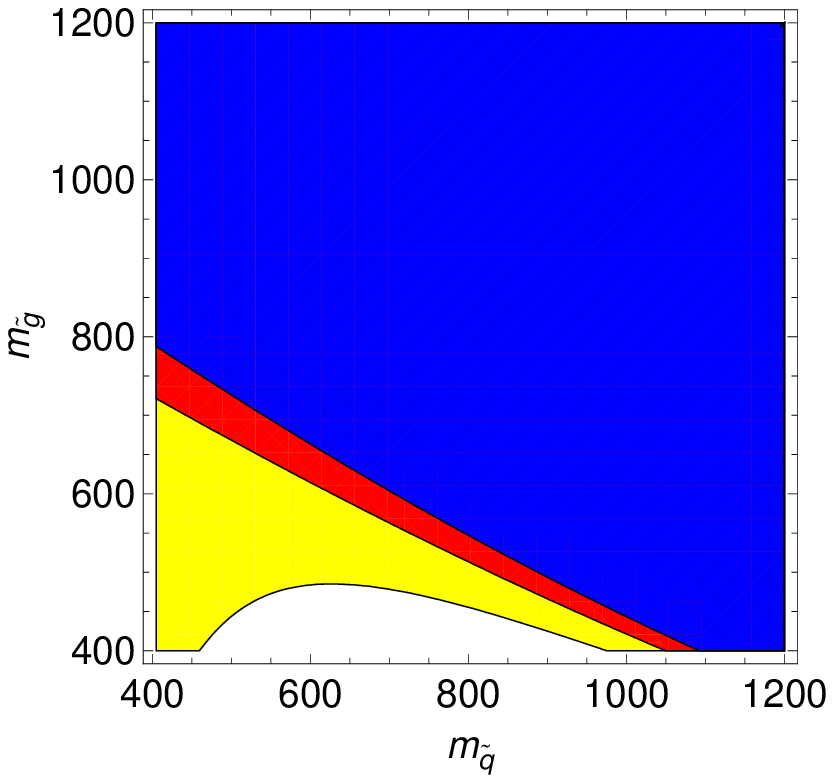}
\caption{Left: Allowed regions in the $m_{\tilde q}-m_{\tilde g}$
  plane. Constraints from $b\to s \gamma$ for different values of 
  $\mu \tan\beta$ assuming that the CKM matrix is generated
  in the down sector. We demand that the calculated branching ratio,
  {with} the SM and {gluino contributions}, lies within the
  $2\,\sigma$ range of the measurement.  Yellow(lightest): allowed region for $\mu
  \tan\beta=30 \tev$, red: $\mu \tan\beta=0 \tev$ and blue(darkest):
  $\mu \tan\beta=-30 \tev$.
  \newline Right: Same as the left plot, but with the weaker
      requirement
that the gluino contribution should not exceed the SM one.
\label{b-s-gamma-allowed}}\hrule
\end{figure*}
The CKM matrix can be generated in the up-sector, in the down-sector
or in both sectors at the same time (see Fig.~\ref{W-Diagramm}).
In the following we concentrate on the two limiting cases in
which either the up-squark or the down-squark sector is flavor-diagonal in
our super-CKM basis. We refer to these two scenarios as ``CKM
generation in the down-sector'' and ``CKM generation in the up
sector'', respectively. For ``CKM generation in the down-sector'' we
obtain from \eq{eq:CKMreno} the following conditions:
\begin{eqnarray}
   \Sigma_{23}^{d\;LR} &=& -m_{b}V_{ts}^*\,\approx\, m_{b}V_{cb}\,,  \nonumber\\
   \Sigma_{13}^{d\;LR} &=& -m_{b}V_{td}^*\,.
\label{CKMdown}
\end{eqnarray}
In principle, the self-energy $\Sigma_{13}^{d\;LR}$ can equivalently be determined from the CKM element $V_{ub}$.
Note, however, concerning $V_{td}$ the situation is a bit more complicated due to the additional doubly flavor-changing contribution $V_{us}\,\dfrac{\Sigma^{d\,LR}_{23}}{m_{b}}$. For ``CKM generation in the up sector'' we have
\begin{eqnarray}
   \Sigma_{23}^{u\;LR} &=& -m_{t}V_{cb}\,\approx\,m_{t}V_{ts}^*  \,,  \nonumber\\
   \Sigma_{13}^{u\;LR} &=& -m_{t}V_{ub}\,.
\label{CKMup}
\end{eqnarray}
For illustration we show in Fig.~\ref{Ad} the size of the $A$-terms
needed to generate the quark masses  and the CKM mixing angles
  according to Eqs.~(\ref{radiative_M}), (\ref{CKMdown}) or
    (\ref{CKMup}). To this end we have set the gluino mass and the
  left-handed and right-handed squark masses to a common value
  $m_{\tilde{q}}$.\medskip

For CKM generation in the up-sector, the required value for $A^{u}_{23}$ is rather large and 
the perturbative vacuum may be unstable \cite{Casas:1996de} or metastable \cite{Park:2010wf} because the scalar 
potential develops a global minimum elsewhere. Arguments based on global 
minima of the classical potential should be taken with the reservation that quantum effects can modify the potential. Close to the perturbative minimum such corrections are calculable and lead to the well-known
loop-corrected effective potential, however we are not aware of a reliable
calculation of a global minimum with non-perturbative methods, which would 
be required in this case. For this reason we do not further consider the bounds from vacuum (meta-)stability which can in principle also be avoided if further heavy particles are added to the MSSM. Note that for CKM generation in the down-sector the vacuum is absolutely stable.
\medskip

\section{Phenomenological consequences for flavor changing processes \label{sec:FCNC}}

Although the B factories have confirmed the CKM mechanism as the
  dominant source of flavor violation with high precision, leaving
little room for new sources of FCNCs in $b\to d$ and $s\to d$
  transitions, we show in this section that radiative generation 
of quark masses and of the CKM matrix still
remains a valid scenario. While flavor-changing transitions among the first
  two generations are CKM-like, this is no longer true once the
third generation is involved, because the $A$-terms are not
diagonal in the same basis as the bare Yukawa couplings. 
It is evident from \eq{cnice} that $A^q_{i3}$ and $A^q_{3i}$ are
non-minimal sources of flavor-violation. In the following we will
concentrate on the two simple limiting cases in which either $A^u$ is
diagonal (in the same basis as $Y^{u(0)}$) and the CKM elements are
generated by the off-diagonal elements of $A^d$, or on the opposite
case in which $A^d$ is diagonal but $A^u$ is not. Even though the
elements $A^{q}_{31,32}$ are not needed for the generation of
the CKM matrix, no symmetry argument requires them to be zero. Note
that it is in principle also possible to generate the fermion masses
with non-holomorphic trilinear terms. Such a scenario (as
proposed in Ref.~\cite{Borzumati:1999sp,Demir:2005ti}) will lead to
additional effects in the Higgs sector \cite{Crivellin:2010er}.

\subsection{CKM generation in the down-sector}

If the CKM matrix is generated in the down sector, the off-diagonal
elements of the squark mass matrix $\Delta^{d\;LR}_{13,23}$ are
determined by the requirement that they generate the observed CKM matrix
via \eq{CKMdown}. Since the off-diagonal elements needed to generate the
CKM matrix in the down-sector are very small ({cf.}\ Fig.~\ref{Ad}),
the mass-insertion approximation excellently reproduces the exact
result. Therefore, we can solve for $\Delta^{d\;LR}_{13,23}$
analytically by using \eq{SQCD-SE-MIA} in \eq{CKMdown}. In the following
we investigate the consequences of the so-determined
$\Delta^{d\;LR}_{13,23}$ on FCNC processes.

\subsubsection{$b\to s\gamma$}

To leading order in the mass insertion approximation, the flavor
off-diagonal elements $\delta^{d\;LR}_{13,23}$
($\delta^{q\;XY}_{ij}=\Delta^{q\;XY}_{ij}/m_{\tilde q}^2$) enter FCNC
processes involving the third generation. Furthermore, also Kaon and D
mixing are affected by diagrams containing the combination
$\delta^{d\;LR}_{13}\times\delta^{d\;RL}_{32}$. Even though Kaon mixing
is very sensitive to NP, especially to new sources of CP violation, the
product $\delta^{d\;LR}_{13}\times\delta^{d\;RL}_{32}$ is too small to
give sizable effects. The contribution to D mixing is even further
suppressed since it is generated by a chargino diagram. However, $b\to
s(d) \gamma$ is very sensitive to $\delta^{d\;LR}_{23}$
($\delta^{d\;LR}_{13}$) since these parameters violate both flavor and chirality.
Even though the relative effect (compared to the SM contribution) in
$b\to s \gamma$ and $b\to d \gamma$ is approximately equal, $b\to s
\gamma$ turns out to be the process which is most sensitive to RFV 
stemming from the down sector \cite{Kagan:1994qg} since it is measured more
precisely than $b\to d \gamma$. The new contributions affect the Wilson
coefficients $C_7$ and $C_8$ of the magnetic and chromomagnetic
operators. The interference of the gluino contribution with the SM contribution is
necessarily constructive. It is important to take into account the
chirally enhanced corrections to the Wilson coefficients discussed in
Ref.~\cite{Crivellin:2009ar}. Due to the inclusion of these effects also
the gluino constraints depend on $\mu$ and $\tan\beta$ (see
Fig.~\ref{b-s-gamma-allowed}).\medskip

\begin{figure*}
\centering
\includegraphics[width=0.65\textwidth]{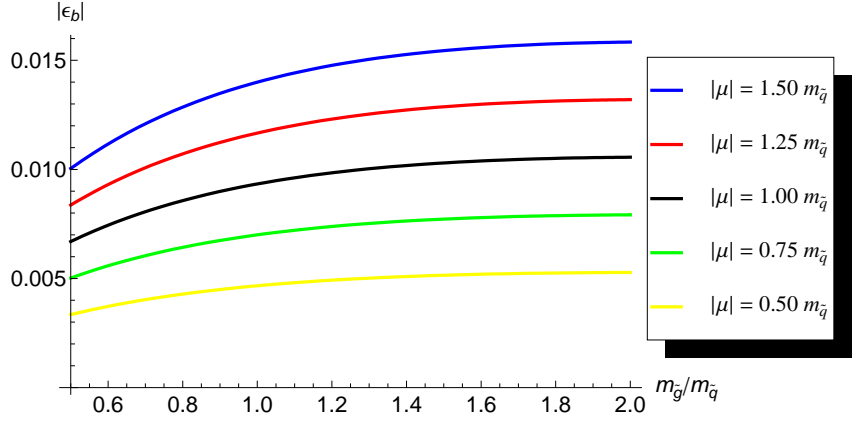}
\caption{$|\epsilon_b|$ as a function of $m_{\tilde g}/m_{\tilde q}$ for different values $|\mu|/m_{\tilde q}$. In the decoupling limit $|\epsilon_b|$ depends only on these two ratios. 
\label{epsilonb}}\hrule
\end{figure*}
The allowed region in the left plot of Fig.~\ref{b-s-gamma-allowed} is
obtained under the assumption that the calculated branching ratio
(including only the SM contribution taken from
Ref.~\cite{Misiak:2006zs,Misiak:2006ab} and the gluino contribution
involving $\delta^{d\;LR}_{23}$) is less than $2\sigma$ away from
  the measured central value of the branching ratio $\rm{Br}\left[
      b\to s \gamma\right]=(3.55 \pm 0.26)\times 10^{-4}$
  \cite{Asner:2010qj}.  However, there are more contributions involving
additional free parameters. First there is the charged Higgs
contribution which depends to a very good approximation only on
$m_{H^+}^2$ and interferes constructively with the SM
\cite{Rizzo:1987km,Besmer:2001cj}. In addition, since no symmetry
argument forbids a non-vanishing term $A^{d}_{32}$ (and therefore
$\delta^{d\;RL}_{23}$) in our model there is another possible
contribution to $C_7^\prime$ and $C_8^\prime$ which can enhance the
branching ratio. However, there is also the chargino contribution which
grows with $\tan\beta$ and can have either sign depending on the product
$\mu A^u_{33}$. Therefore, this contribution can interfere destructively
with the SM, the gluino, and the charged Higgs contribution. In total
the branching ratio can be in agreement with experiment. This is
possible for a wide range of parameters, however some degree of
fine-tuning is necessary. In order to avoid very large cancellations,
one can demand that none of the various NP contributions contribution
should exceed the SM one. Under this assumption the allowed region in
the right plot of Fig.~\ref{b-s-gamma-allowed} is obtained.

\subsubsection{Non-decoupling {Higgs-mediated} effects}

At moderate-to-large $\tan\beta$ Higgs-mediated effects become
important. These effects are non-decoupling; this means that they
do not vanish for heavy SUSY masses but only decouple like
$1/M_{\rm{H}}^2$ (for large $\tan\beta$ the CP-even Higgs, the charged
Higgs and the CP-odd Higgs have approximately equal masses $m_{H}\equiv
m_{H^0}\approx m_{A^0}\approx m_{H^+}$). As shown recently
\cite{Crivellin:2010er,Crivellin:2011jt}, if one consistently includes
all chirally enhanced effects into the calculation of the effective
Higgs vertices, the trilinear $A$-terms induce effective flavor-changing
neutral Higgs couplings. These effects are important in our model
already for moderate $\tan\beta$ since we necessarily have flavor
off-diagonal $A$-terms in order to generate the CKM matrix.\medskip

Following Refs.~\cite{Crivellin:2010er,Crivellin:2011jt} and neglecting
terms proportional to $\cos\beta$, a Feynman rule for the effective
neutral Higgs coupling mediating $b$-$s$ transitions (The corresponding
formula for $b$-$d$ transitions are simply obtained by exchanging $s$
and $d$) is given by
\begin{equation}
i\left( {\Gamma _{sb}^{H_k^0\,LR} P_R  + \Gamma _{sb}^{H_k^0\,RL} P_L } \right)
\end{equation}
with
\begin{equation}
 \Gamma _{sb}^{H_k^0\,LR} \, =\, \Gamma _{bs}^{H_k^0\,RL*} \, =\, x_d^{k}\, \dfrac{1}{{v_d }}\, \dfrac{\Sigma _{23}^{d\;LR}}{{m_{b} }}\,\Sigma _{33}^{\prime d\;LR}  \,.
\end{equation}
Here $H_k^0$ denotes the three physical neutral Higgs bosons: the heavy
CP-even Higgs $H_1^0=H^0$, the light CP even Higgs $H_2^0=h^0$ and the
CP-odd Higgs $H_3^0=A^0$. For $H^0_k=(H^0,h^0,A^0)$ the coefficients
$x_d^{k}$ are given by
\begin{equation}  
  x_d^{k}  = \left( {\dfrac{{ - 1}}{{\sqrt 2 }}\cos \left( \alpha  \right),\dfrac{1}{{\sqrt 2 }}\sin \left( \alpha  \right),\dfrac{i}{{\sqrt 2 }}\sin \left( \beta  \right)} \right)\,.
\end{equation}
Furthermore, $\Sigma _{33}^{\prime d\;LR}$ denotes the non-holomorphic
part of the self-energy which is proportional to the $\mu$\,-\,term originating from
$\Delta^{d\,LR}_{33}$ in \eq{DeltaLR}. It is given by
\begin{equation}
  \Sigma _{33}^{\prime d\;LR}  = m_b 
  \dfrac{{\varepsilon _b \tan \beta  }}{{1 + \varepsilon _b\tan \beta  }}
\end{equation}
with
\begin{equation}
\varepsilon _b  = -\dfrac{2\alpha_s}{{3\pi }} m_{\tilde g} \mu\, C_0 \left( {m_{\tilde g}^2 ,m_{\tilde b_1 }^2 ,m_{\tilde b_2 }^2 } \right).
\label{eq:epsilonb}
\end{equation}
The loop function $C_0$ can for example be found in
Ref.~\cite{Hofer:2009xb}.  Due to the mass eigenstates in the
loop-function, \eq{eq:epsilonb} is also valid beyond the decoupling
limit in the absence of flavor violation {\cite{Carena:1999py,Hofer:2009xb}}. In the
decoupling limit (which is an excellent approximation
\cite{Crivellin:2010er}) with degenerate diagonal squark mass terms
$m_{\tilde q}^2$, $\varepsilon _b $ is only a function of the two ratios
$m_{\tilde g}/m_{\tilde q}$ and $\mu/m_{\tilde q}$. We see from
Fig.~\ref{epsilonb} that typical values for $|\varepsilon _b |$ range from
0.005 to 0.01.\medskip
\begin{figure}
\centering
\includegraphics[width=0.45\textwidth]{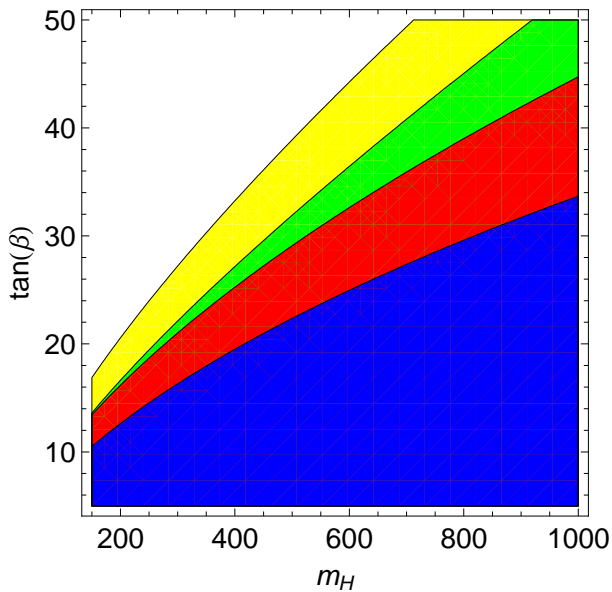}
\caption{Allowed region in the $m_{H}$--$\tan\beta$ plane for different
  values of $\epsilon_b$ from {$\rm{Br}[B_s\to\mu^+\mu^-]\leq 3.6\cdot
  10^{-8}[95\% CL] $}. 
  Yellow: $\epsilon_b=0.005$, green: $\epsilon_b=0.01$, 
  red: $\epsilon_b=-0.005$, 
  blue: $\epsilon_b=-0.01$ (light to dark).
  \label{Bs-mumu}}\hrule
\end{figure}

With $\Sigma _{23}^{d\;LR}$ being fixed by \eq{CKMdown}, $\Gamma _{sb}^{H_k^0\,LR}$ 
becomes
\begin{equation}
\Gamma _{sb}^{H_k^0 \,LR}  \,=\, -x_d^k\, \dfrac{{V_{ts}^*}}{{v_d }}\,\Sigma_{33}^{\prime d\,LR}\,.
\end{equation}
This effective Higgs coupling induces a contribution to
$B_s\to\mu^+\mu^-$ and therefore gets constrained from this process.
Fig.~\ref{Bs-mumu} shows the allowed region compatible with the
  experimental bound of Refs.~\cite{Asner:2010qj,:2007kv,Abazov:2010fs} in the
  $\tan\beta$--$m_{H}$ plane, where $m_{H}$ is the heavy Higgs
    mass.  We see that $\rm{Br}[B_s\to\mu^+\mu^-]$ can be
significantly enhanced even for moderate values of $\tan\beta$.\medskip
\begin{figure}
\centering
\includegraphics[width=0.45\textwidth]{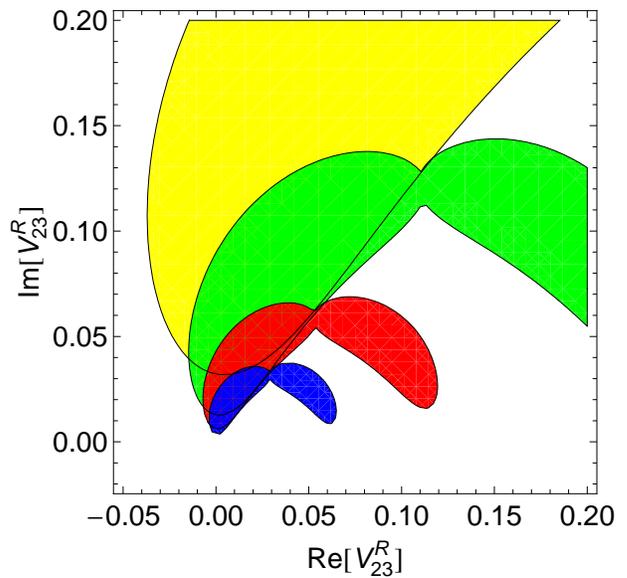}
\caption{Allowed region (95\% confidence level) in the  
$\rm{Re}[V^R_{23}]$--$\rm{Im}[V^R_{23}]$ plane for $\epsilon_b=0.0075$, 
$m_H=400\rm{GeV}$ and different values of $\tan\beta$ from \bbms. 
Yellow: $\tan\beta=11$, green: $\tan\beta=14$, red:
$\tan\beta=17$, blue: $\tan\beta=20$ (light to dark).
The peculiar shape stems from the fact that the data permit two 
solutions for the \bbms\ phase. The solution with the smaller 
values of $\real[V^R_{23}]$ corresponds to the solution 
closer to the SM value.\label{Bs-mixing}}\hrule
\end{figure}

\begin{figure*}
\centering
\includegraphics[width=0.44\textwidth]{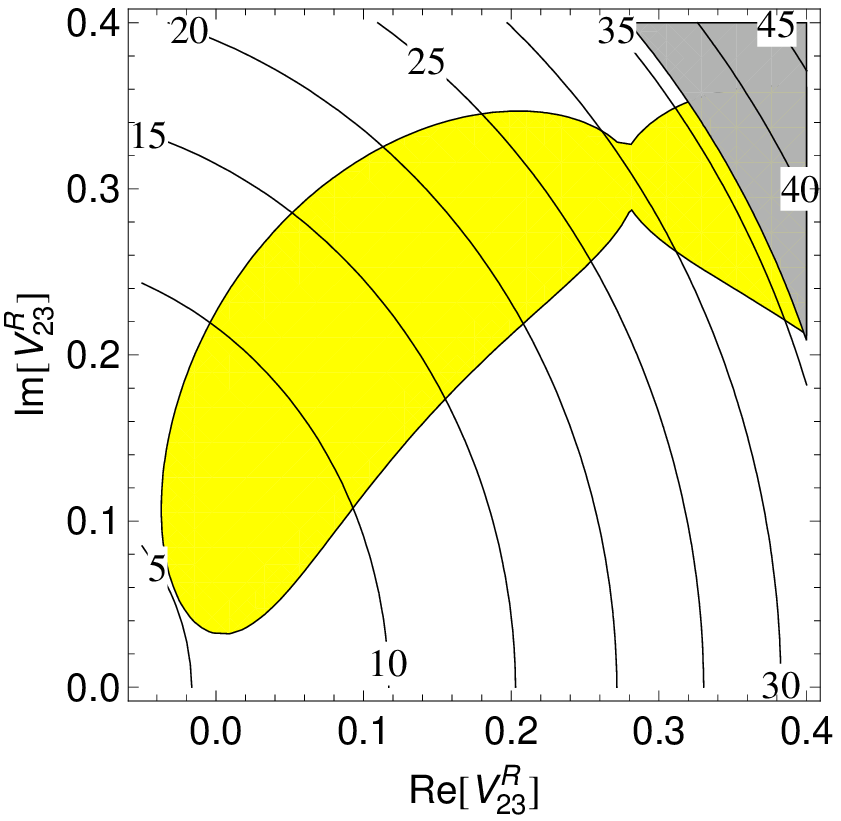}\hspace{0.08\textwidth}
\includegraphics[width=0.45\textwidth]{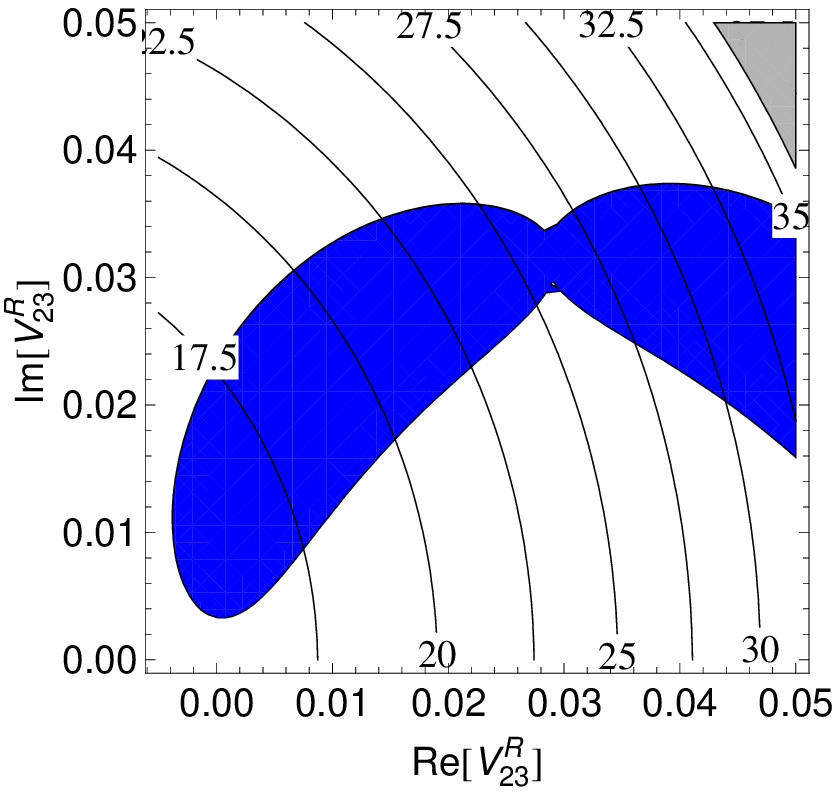}
\caption{
Correlations between $B_s\to \mu^+\mu^-$ and \bbms\ for 
$\epsilon_b=0.0075$, $m_H=400\rm{GeV}$. 
Left plot, yellow: Allowed region from \bbms\ (95\% confidence
level) for $\tan\beta=11$. The contour-lines show $\rm{Br}[B_s\to
\mu^+\mu^-]\times10^9$. The grey area at the right side is excluded by the
Tevatron bound on $\rm{Br}[B_s\to\mu^+\mu^-]$. 
Right plot, blue: same for $\tan\beta=20$.\label{Bs-mixing2}}\hrule
\end{figure*}

As already stated, no symmetry argument forbids a non-vanishing value of
$\Delta^{d\;LR}_{32}$. Such a term would generate a chirally enhanced
self-energy $\Sigma _{32}^{d\;LR}$ which would in turn lead to an
effective Higgs coupling $\Gamma _{bs}^{H_k^0 \,LR}$. In this case we
also get a contribution to \bbms. This additional
contribution might explain the observed deviation of the measured
\bbms\ phase from the SM expectation
{\cite{Abazov:2010hj,Abazov:2010hv,taggedphaseCDF_3,taggedphaseD0_2,Lenz:2010gu}.}
{The SM contributions to the width difference appearing in the
  angular analysis of the $B_s \to J/\psi \phi$ data and to the CP
  asymmetry in flavor-specific decays (equivalent to the dimuon
  asymmetry) have been calculated in
  Refs.~\cite{Beneke:1998sy,Beneke:2003az,Lenz:2006hd}.}  In
{Fig.~\ref{Bs-mixing}} we show the regions in parameter space which
are consistent with \bbms\ at the 95\% confidence
level {using the fit result of Ref.~\cite{Lenz:2010gu}}.  In order to
simplify the notation and to allow for an easy comparison with the size
of $\Sigma_{23}^{d\;LR}=-m_{b}V_{ts}^*$ we have defined the quantity
\begin{equation}
	V^{R}_{23}=-V^{R\,*}_{32}=\Sigma _{23}^{d\;RL}/m_b.
\end{equation}
In {Fig.~\ref{Bs-mixing2}} we show the {correlation} between
\bbms\ and $B_s\to \mu^+\mu^-$ for $m_H=400
  \rm{GeV}$, $\epsilon_b=0.0075$ and two values of $\tan\beta$.  Note
that the region in parameter space which can explain the phase in \bbms\
is well compatible with the current limits on
$\rm{Br}[B_s\to\mu^+\mu^-]$.  {Moreover, if the hints for a sizable
  new-physics contribution to \bbms\ persist, $B_s\to\mu^+\mu^-$ will
  necessarily be enhanced.}  LHCb will be able to probe this correlation
in the near future.  {The global analysis in Ref.~\cite{Lenz:2010gu}
  also finds a preference for a new CP phase $\phi_d^\Delta \approx
  -13^\circ$ in \bbmd\ adding to the SM phase of $2\beta$. Since a
  smaller new-physics contribution than in the $B_s$ system is needed,
  it is easy to accomodate this with the free parameter
  $\delta^{d\;RL}_{13}$.  (The experimental bound on the relevant ratio
  $\rm{Br}[B_d\to\mu^+\mu^-]/|V_{td}|^2$ is weaker than on
  $\rm{Br}[B_s\to\mu^+\mu^-]/|V_{ts}|^2$.)}  In summary, if the CKM
matrix is generated in the down-sector, sizable Higgs-induced effects
are generated (even in the decoupling limit) which can enhance $B_s\to
\mu^+\mu^-$ and can {accommodate} the observed evidence for new
  CP-violating physics in \bbms.\medskip

Finally we want to discuss the correlation between the bounds from $b\to
s \gamma$ and the non-decoupling constraints. The constraints on the
SUSY-masses shown in Fig.~\ref{b-s-gamma-allowed} are weakened for positive $\mu$ 
and large values of $\tan\beta$. In
addition, at large $\tan\beta$ also the chargino contribution to $b\to s
\gamma$ becomes important and can interfere destructively with the
gluino and the SM contribution lowering the bound on the SUSY masses.
Thus the constraint on the SUSY masses can only be lowered for large
values of $\tan\beta$ which then leads to sizable contributions to
$B_s\to\mu^+\mu^-$ (see Fig.~\ref{Bs-mumu}). Since the bounds from
$B_s\to\mu^+\mu^-$ do not decouple with the SUSY scale but only with the
Higgs mass they do not vanish for heavy SUSY masses like the constraints
from $b\to s \gamma$.\medskip

\subsection{CKM generation in the up-sector}

If the CKM matrix is generated in the up-sector, the off-diagonal
elements of the squark mass matrix $\Delta^{u\,LR}_{13,23}$ are
determined by \eq{CKMup}. Due to the large top mass these off-diagonal
elements are much larger than in the case of CKM generation in the down
sector. Therefore, already the
requirement that the lighter stop mass does not violate the bounds from
direct searches requires the diagonal elements of the squark mass matrix
to be heavier than approximately $(700\,\rm{GeV})^2$. Furthermore, the
mass-insertion approximation (MIA) does not necessarily hold for such
large off-diagonal elements. One cannot solve {the exact expressions}
analytically for $\Delta^{u\;LR}_{13,23}$ but rather has to determine
these elements numerically. However, for squark masses above $700\,
\rm{GeV}$ {we find} the off-diagonal elements determined {in} MIA
{larger} than the ones obtained by {exact} diagonalization {by just ten
  percent or less.}  Therefore, it is still possible to rely on MIA for
a qualitative understanding of the flavor structure.
\medskip

If the CKM matrix is generated in the up-sector, one naively {expects
  the} chargino contributions to $b\to s \gamma$, $b\to d \gamma$ and
$B_{s,d}\!-\overline{\!B}_{s,d}$ mixing {to give relevant bounds on
  $\delta^{u\;LR}_{13,23}$}. However, $B_{s,d}\!-\overline{\!B}_{s,d}$ mixing
does not give useful constraints on $\delta^{u\;LR}_{13,23}$ and
{$\textrm{Br}[b\to s \gamma,d\gamma]$} also heavily depends on $\mu$ and
$\tan\beta$.  Furthermore, we again have to take into account the
chirally enhanced effects by using the effective chargino vertices given
in Ref.~\cite{Crivellin:2008mq}.  In the present case of RFV 
in the up-sector, these effective vertices read
\begin{eqnarray}
 \Gamma _{d_i \tilde u_s }^{\tilde \chi _k^ \pm  L}  &=& \sum\limits_{j = 1}^3 {V^{(0)}_{C\,ji} \left( Y^{t(0)} \delta _{j3}\,{V_{k2}^{\tilde \chi ^ \pm  *}  W_{j + 3,s}^{\tilde u*}  - g_2 V_{k1}^{\tilde \chi ^ \pm  *} W_{js}^{\tilde u*} } \right)},  \nonumber\\ 
 \Gamma _{d_i \tilde u_s }^{\tilde \chi _k^ \pm  R}  &=& Y^{b(0) } \delta _{i3}\, U_{k2}^{\tilde \chi ^ \pm  }  \sum\limits_{j = 1}^3 {V^{(0)}_{C\,ji} W_{js}^{\tilde u*} }. 
\end{eqnarray}
Note that it is the bare CKM matrix $V^{(0)}_{C}$ (with vanishing
elements connecting the third with the first two generations) and not
the physical CKM matrix $V$ which appears in these couplings to external
down-type quarks. This is easy to understand since the physical CKM
matrix $V$ is generated in the up-sector meaning that the down-type
quarks are not rotated by flavor-changing self-energies.  Note
further that the Yukawa couplings of the quarks of the first two
generations are zero in our scenario of RFV. \medskip

\begin{figure}
\centering
\includegraphics[width=0.45\textwidth]{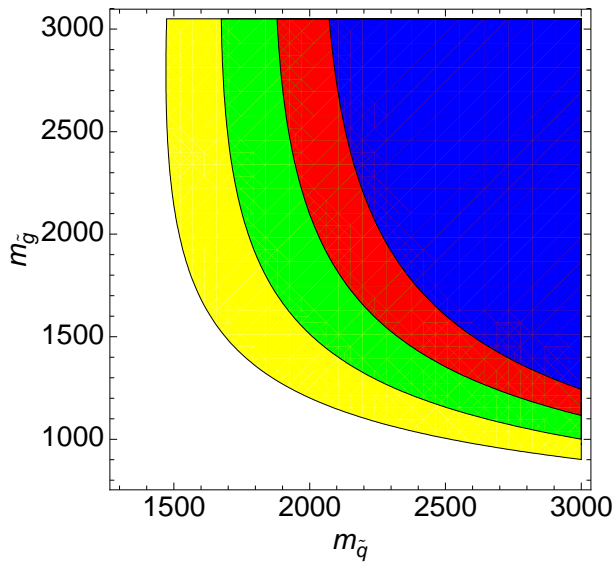}
\caption{Allowed regions in the $m_{\tilde q}-m_{\tilde g}$ plane. Constraints from \kkm\ for different values of $M_2$ assuming that the CKM matrix is generated in the up sector. Yellow(lightest): $M_2=1000 \gev$, green: $M_2=750\gev$, red: $M_2=500\gev$ and blue(darkest): $M_2=250\gev$.
\label{K-allowed}}\hrule
\end{figure}

While $b\to s\gamma$, $b\to d \gamma$ and $B_{s,d}\!-\overline{\!B}_{s,d}$ mixing
  do not give severe constraints on our model of RFV, there exist other
(less obvious) contributions to \kk and {\ddm} which must be taken
  into account. An effective element $\delta^{u\;LL}_{12\;\rm{eff}}$ is
induced through the double mass insertion $\delta^{u\;LR}_{13}\times
\delta^{u\;LR*}_{23}$. Note that this element is proportional to two
powers of an electroweak vev and is therefore not subjected to the
$SU(2)$ relation which relates $\delta^{u\;LL}_{ij}$ to
$\delta^{d\;LL}_{ij}$. Therefore, on the one hand only chargino diagrams
contribute to \kkm\ while on the other hand only gluino diagrams
contributes to \ddm\ mixing. However, in the case of \kkm\ we have very
precise experimental information on CP violation, the corresponding
quantity $\epsilon_K$ is well understood in the SM
\cite{Herrlich:1995hh,Herrlich:1996vf,Brod:2010mj}. Since moreover
$\delta^{u\;LL}_{12\;\rm{eff}}$ carries the CKM phase $\gamma$ (because
it is proportional to $\delta^{u\;LR}_{13}$ which generates $V_{ub}$),
the constraint from $\epsilon_K$ turns out to be stronger than the
constraints from D mixing \cite{Crivellin:2010ys}.  The allowed regions
in the $m_{\tilde q}-m_{\tilde g}$ plane for different values of $M_2$
are shown in Fig.~\ref{K-allowed}.  Note that the constraints are nearly
independent of $\mu$ and $\tan\beta$ since the quark-squark coupling
involves the gaugino component of the charginos.\medskip

\begin{figure*}
\centering
\includegraphics[width=0.8\textwidth]{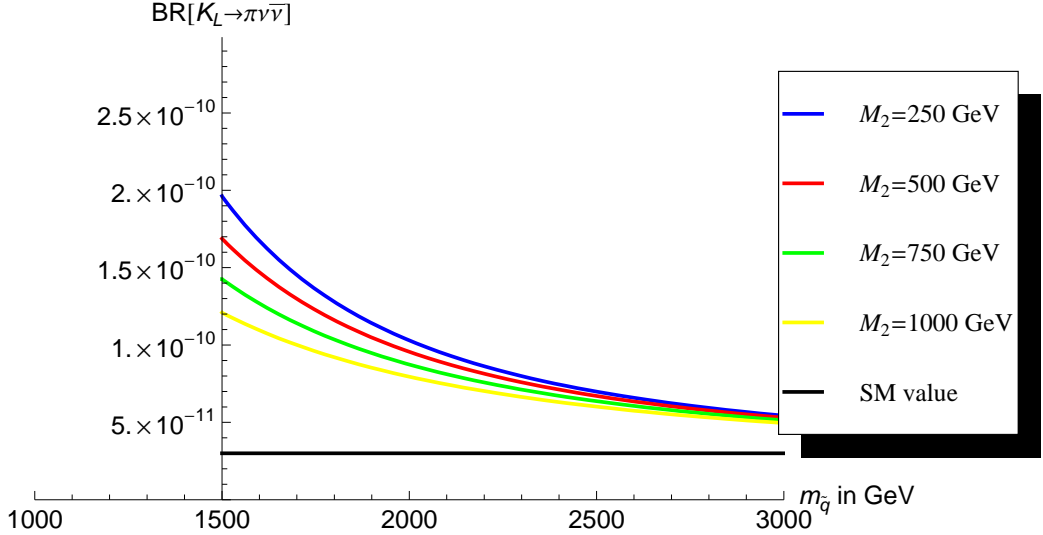}
\caption{Predicted branching ratio for the rare Kaon decay $K_L\to\pi\nu\overline{\nu}$ assuming that the CKM matrix is generated in the up-sector for $m_{\tilde{q}}=m_{\tilde{g}}$. The branching ratio is enhanced due to a constructive interference with the SM contribution.
  \label{Klongtopinunu}}
\hrule
\end{figure*}

\begin{figure*}
\centering
\includegraphics[width=0.8\textwidth]{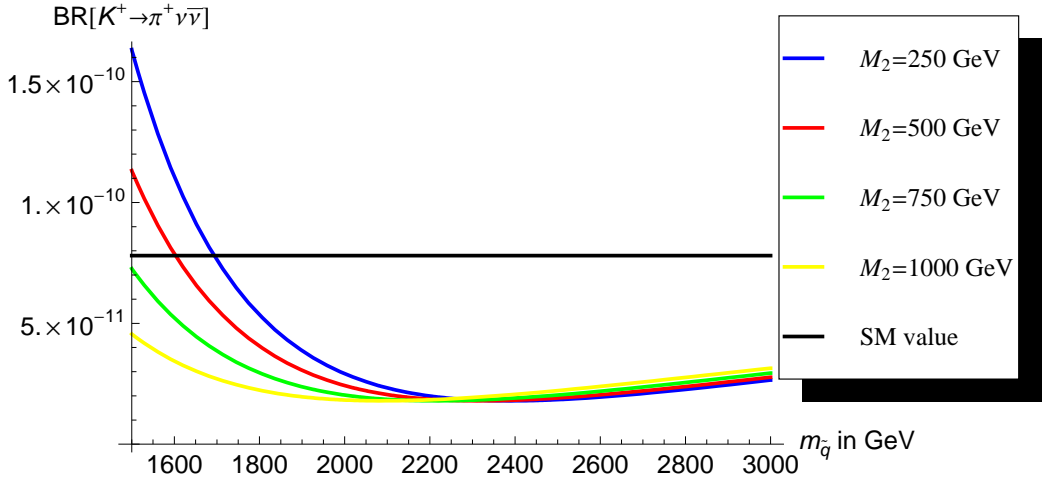}
\caption{Predicted branching ratio for the rare Kaon decay $K^+\to\pi^+\nu\overline{\nu}$ assuming that the CKM matrix is generated in the up-sector for $m_{\tilde{q}}=m_{\tilde{g}}$. The branching ratio is enhanced for light SUSY masses but suppressed if the scale of SUSY breaking is higher.
\label{Kplustopinunu}}\hrule
\end{figure*}

Another process which is sensitive to the combination
$\delta^{u\;LR}_{13}\times \delta^{u\;LR*}_{23}$ via chargino loops is
$K\to\pi\nu\nu$
\cite{Nir:1997tf,Buras:1999da,Colangelo:1999kr,Buras:2004qb}. Even
though, at present, this process does not give useful bounds, but
NA62 results will change this situation in the future.
Fig.~\ref{Klongtopinunu} and Fig.~\ref{Kplustopinunu} {show} the
predicted branching ratios for $K_L\to\pi\nu\overline{\nu}$ and
$K^+\to\pi^+\nu\overline{\nu}$. In a wide range of squark and gluino
  masses both quantities largely deviate from the SM predictions
  \cite{Buras:2005gr,Buras:2006gb,Brod:2010hi}, the effect on the
  charged mode should be detectable at NA62.  Our values for
  $\rm{Br}[K_L\to\pi\nu\overline{\nu}]$ plotted in
  Fig.~\ref{Klongtopinunu} overlap with the region probed by the KOTO
  experiment at JPARC.  Note that the branching ratios are again to a
very good approximation independent of $\mu$ and $\tan\beta$.\medskip

In principle $\delta^{u\;LR}_{23}$ also contributes to $B\to
K\nu\overline{\nu}$ via a Z penguin (and at the same time also to
$B_s\to \mu^+\mu^-$ which is strongly correlated to $B\to
K\nu\overline{\nu}$ in the MSSM at low $\tan\beta$)
\cite{Bertolini:1990if,Yamada:2007me,Altmannshofer:2009ma}. Even though
the branching ratios are slightly enhanced, they also depend on
$A^u_{33}$, $\mu$ and $\tan\beta$. Furthermore, $B\to
K\nu\overline{\nu}$ is also correlated to $b\to s\gamma$ which forbids
large effects \cite{Yamada:2007me}.
Of course also in the up-sector no symmetry argument forbids non-zero
elements $\delta^{u\;LR}_{31,32,33}$. While, {as already discussed,}
$\delta^{u\;LR}_{33}$ affects $b\to s\gamma$ and $B\to
K\nu\overline{\nu}$, the elements $\delta^{u\;LR}_{31,32}$ are rather
{unconstrained} from FCNC processes, since they enter these processes
only in combination with small quark masses and small chargino mixing.
Their mere effect is to correct the eigenvalues {of the up-type
  squark mass matrix.}  However, as shown in
Ref.~\cite{Crivellin:2009sd}, they can induce a sizable right-handed
$W$-coupling if at the same time also $\delta^{d\;LR}_{33}$ is large.

\section{Conclusions}

Radiative generation of light fermion masses and CKM mixing angles is
{an appealing} and very predictive concept. Within the MSSM this
approach can solve the SUSY CP and flavor problems
{\cite{Borzumati:1999sp,Crivellin:2008mq}}. In this article we have
studied a model with radiative flavor violation (RFV) using the
trilinear terms as spurions breaking flavor symmetries
\cite{Crivellin:2008mq} and we have analyzed its phenomenological
implications on FCNC processes.  Keeping the third-generation fermion
Yukawa couplings, the CKM matrix can either be induced in the up or in
the down sector. (In principle also a mixed scenario is possible,
however, we did not further investigate this possibility.) If the CKM
matrix is generated in the up-sector, Kaon mixing severely constrains
the allowed values of $m_{\tilde g}$ and $m_{\tilde q}$ (see
Fig.~\ref{K-allowed}). However, the rare Kaon decays $K\to
\pi\nu\nu$ can still receive sizable contributions. If the other
possibility is realized and the CKM matrix is generated in the down
sector, $b\to s\gamma$ restricts the allowed range for the SUSY masses.
{Fig.~\ref{b-s-gamma-allowed} shows our results which} take into account
the chirally enhanced correction discussed in
Refs.~\cite{Crivellin:2009ar,Crivellin:2011jt}. In the case of CKM
generation in the down sector $B_s\to\mu^+\mu^-$ receives sizable
contributions (even in the decoupling limit) {from} flavor-changing
effective Higgs couplings, {which are already sizable} at moderate
$\tan\beta$ \cite{Crivellin:2010er}. If in addition $\delta
_{23}^{d\;RL}\neq0$, \bbms\ is affected by double Higgs penguins as
well. In this way the \bbms\ phase which disagrees with the SM
expectation \cite{Lenz:2010gu} can be explained.\medskip

{Having shown that all CKM elements and small quark masses can be
  generated from loop diagrams while simultaneously obeying all FCNC
  constraints, we conclude that the MSSM with RFV is a viable
  alternative to the popular MFV variant of the MSSM. As opposed to the 
  MFV-MSSM our RFV model is capable to explain the large \bbms\
  phase favored by current data.}

\section*{Note added}\vspace{-0.02\textheight}
After this paper was completed, the anomaly of the like-sign dimuon charge asymmetry measured by the D0 experiment increased from $3.2\sigma$ to $3.9\sigma$ \cite{Abazov:2011yk}. Furthermore, the CDF experiment has reported hints for an enhanced $B_s\to \mu^+ \mu^-$ branching ratio \cite{Collaboration:2011fi}. This nicely complies with the fact that in the scenario with down-sector RFV a NP contribution to \bbs mixing favors an enhanced $B_s\to \mu^+\mu^-$ branching ratio (see Fig.~7). However, the situation is inconclusive, because LHCb \cite{LHCb} and CMS \cite{CMS} do not confirm these findings.
Their 95\% CL upper bounds $\rm{Br}[B_s\to\mu^+\mu^-]=1.5\times 10^{-8}$ and 
$\rm{Br}[B_s\to\mu^+\mu^-]=1.9\times 10^{-8}$, respectively, are both compatible with 
the two-sided bound of CDF and with the SM expectation.

\bigskip

{\it Acknowledgments.}---{We are grateful to Miko\l aj Misiak for a
  critical reading of the manuscript.}  A.C.\ thanks David Straub for
proposing Fig.~\ref{Bs-mixing2}. This work is supported by BMBF grants
05H09WWE and 05H09VKF and by the EU Contract No.~MRTN-CT-2006-035482,
\lq\lq FLAVIAnet''. A.C.\ acknowledges the financial support by the State
of Baden-W\"urttemberg through \emph{Strukturiertes Promotionskolleg
  Elementarteilchenphysik und Astroteilchenphysik} and by the Swiss National Foundation.  The Albert Einstein Center
for Fundamental Physics is supported by the ``Innovations- und
Kooperationsprojekt C-13 of the Schweizerische Universit\"atskonferenz
SUK/CRUS''. L.H.\ and D.S.\ acknowledge the support by Ev.\ Studienwerk
Villigst and by Cusanuswerk, respectively. L.H. has been
further supported by the Helmholtz Alliance "Physics at the Terascale". 

\bibliography{radiative-generation}

\end{document}